# Flat Maps that improve on the Winkel Tripel

J. Richard Gott III[1], David M. Goldberg[2], and Robert J. Vanderbei[3]

1. Department of Astrophysical Sciences, Princeton University, Princeton, NJ, USA

2. Department of Physics, Drexel University, Philadelphia, PA, USA

3. Department of Operations Research, Princeton University, Princeton, NJ, USA

## ABSTRACT

Goldberg & Gott (2008) developed six error measures to rate flat map projections on their verisimilitude to the sphere: Isotropy, Area, Flexion, Skewness, Distances, and Boundary Cuts. The first two depend on the metric of the projection, the next two on its first derivatives. By these criteria, the Winkel Tripel (used by National Geographic for world maps) was the best scoring of all the known projections with a sum of squares of the six errors of 4.563, normalized relative to the Equirectangular in each error term. We present here a useful Gott-Wagner variant with a slightly better error score of only 4.497. We also present a radically new class of flat double-sided maps (like phonograph records) which have correct topology and vastly improved error scores: 0.881 for the azimuthal equidistant version. We believe it is the most accurate flat map of Earth yet. We also show maps of other solar system objects and sky maps.

## Introduction

In terms of shape and area distortions, the Winkel-Tripel projection is hard to beat. Goldberg & Gott (2008) introduced a metric for comparing the fidelity of whole world map projections: Isotropy, Area, Flexion (the apparent bending of great circle routes on the map), Skewness (lopsidedness), Distances, and Boundary Cuts. By evaluating the fidelity metric for a wide range of popular (and occasionally obscure) map projections, it was found that the Winkel Tripel scored best among known map projections for whole planet maps.

But is the Winkel Tripel it the best *possible* map? Not necessarily. In this work, we explore several new projections, including a "conventional" map (the Gott-Wagener) which outperforms the Winkel-Tripel, along with a family of double-sided maps which perform significantly better.

The first two error terms are described by "Tissot Indicatrices" (Tissot 1859): $I$ for isotropy and $A$ for area. A Tissot ellipse is the image of an infinitesimal circle on the globe on the map, with semi-major and -minor axes, $a$ and $b$. $I$ is the R.M.S (root mean square) of $\ln(a_i/b_i)$ over random points $P_i$ sampled uniformly on the globe. *A* is the R.M.S. value of $\ln(a_i b_i /<a_i b_i>)$, where $a_i b_i$ is proportional to the area of the Tissot ellipse at a particular point "i" and $<a_i b_i>$ is proportional to the average area of all Tissot ellipses, with points "i" sampled uniformly over the globe. Fractional errors here are important, as a 4 to 1 error in area (or isotropy) is twice as bad as a 2 to 1 error because it is like making a 2 to 1 error twice.

Flexion is bending on a flat map of a geodesic on the globe. For example, the geodesic great circle route from New York to Tokyo on a Mercator map is a curved line. A flexion of F = 1 is a bending of 1 degree per degree of travel on the geodesic. On a complete stereographic projection of the globe onto a plane (missing the South Pole) a random geodesic is shown as a circle, so averaging over all geodesics, this gives this projection a value of F = 1. S is skewness, or lopsidedness. On a Mercator map the geographic center of North America is in Canada rather than in North Dakota where it should be: Canada is too big and Mexico is too small, making North America lopsided to the North. Imagine a truck traveling on a geodesic path at constant speed north along a meridian in North America. Flexion is a sideways acceleration of the truck image on the map, the driver would be turning his wheel. Since his path north is straight on the Mercator map in this case: there is no flexion on his path. But because the scale is changing, the truck is accelerating, increasing its speed, as it goes north. This acceleration is the measure of skewness: forward or backward acceleration on the map instead of sideways acceleration as would be the case for flexion. Skewness and Flexion depend on the first derivatives of the map metric. One averages the absolute value of the bending over all geodesics as they emanate from random points on the globe. The flexion error is proportional to the bending. A bending of 2° is twice as bad as a bending of 1° because it is equivalent to making a 1° bending error twice in a row. Skewness acceleration is treated similarly.

On conformal maps like the complete stereographic, flexion and skewness errors are equal, in that case being F = S = 1. Kerkovits (2018) has extended these studies with new methods of calculation and visualization of flexion and skewness.

Distance errors, *D*, are R.M.S. $\ln(d_{map,i,j}/d_{globe,i,j})$, where we average over random pairs of points $P_i$ and $P_j$ on the globe. An overestimated distance (measured by a string connecting the two points on a map) by a factor of 2 is just as bad as an underestimated distance by a factor of 2. Finally, the boundary cut error, *B*, is the length of the boundary cut in degrees divided by 720°. Thus, the 180° cut on the Winkel Tripel has an error of B = 0.25 A boundary cut of 360° is twice as bad as a boundary cut of 180° because it is like making two boundary cuts of 180° in a row. Table I has error values for some standard map projections.

## Table 1: Some Standard Projections

|                  | I     | A     | F     | S     | D     | B    |
|------------------|-------|-------|-------|-------|-------|------|
| Equirectangular  | 0.51  | 0.41  | 0.64  | 0.60  | 0.449 | 0.25 |
| Mercator         | 0     | 0.84  | 0.64  | 0.64  | 0.440 | 0.25 |
| Lagrange         | 0     | 0.73  | 0.53  | 0.53  | 0.432 | 0.25 |
| Briesemeister    | 0.79  | 0     | 0.81  | 0.42  | 0.372 | 0.25 |
| Eckert IV        | 0.70  | 0     | 0.75  | 0.55  | 0.390 | 0.25 |
| Winkel-Tripel    | 0.49  | 0.22  | 0.74  | 0.34  | 0.374 | 0.25 |
| Gott-Wagener     | 0.395 | 0.319 | 0.682 | 0.367 | 0.397 | 0.25 |

(The Gott-Wagner projection appears in Figure 1 and will be discussed in the next section.) The combination of the six error terms depends on the choice of normalizing projection (the one which gets a score of 1 for each term). Following precedent, see Laskowski (1997a,b), Goldberg and Gott (2008) use the first and simplest map projection in *Flattening the Earth* (Snyder, 1993): the equirectangular (x = longitude, y = latitude).

**Normalized squared errors ($\sigma^2$)**

**Some Standard Projections**

| $\sigma^2$       | I     | A     | F     | S     | D     | B | $\Sigma\sigma^2$ |
|------------------|-------|-------|-------|-------|-------|---|------------------|
| Equirectangular  | 1     | 1     | 1     | 1     | 1     | 1 | 6                |
| Mercator         | 0     | 4.203 | 1.0   | 1.145 | 0.960 | 1 | 8.308            |
| Lagrange         | 0     | 3.168 | 0.689 | 0.774 | 0.925 | 1 | 6.556            |
| Briesemeister    | 2.403 | 0     | 1.613 | 0.490 | 0.687 | 1 | 6.193            |
| Eckert IV        | 1.877 | 0     | 1.369 | 0.846 | 0.755 | 1 | 5.8474           |
| Winkel-Tripel    | 0.923 | 0.288 | 1.337 | 0.321 | 0.694 | 1 | 4.563            |
| Gott-Wagener     | 0.600 | 0.605 | 1.136 | 0.374 | 0.782 | 1 | 4.497            |

The metric gives a normalized value of 6.0 for sum of the squares of the 6 errors for the equirectangular projection, with lower scores corresponding to better maps. Minimizing sums of squares of errors is standard procedure for model fitting dating back to Gauss. A perfect map (a globe) would have a score of 0. The Winkel Tripel (used by National Geographic for world maps) was the best scoring of all the known projections with a normalized sum of squares of the six errors of 4.563. See Figure 1. The goal of the Winkel Tripel was to minimize isotropy, area, and distance errors. Figure 1 also shows the Gott-Wagner projection which is slightly better with a normalized sum of squares of errors of 4.497. We will describe the Gott-Wagner projection in the next section.

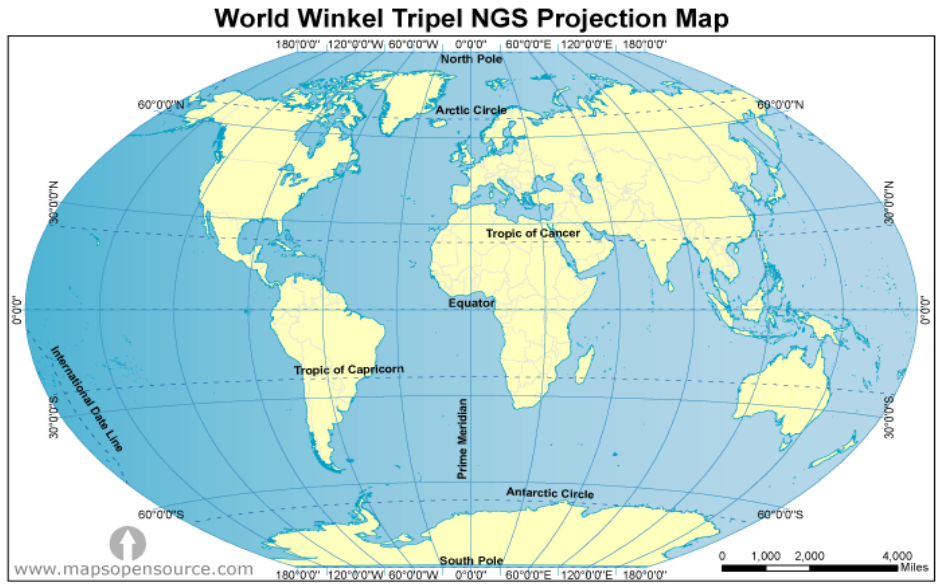

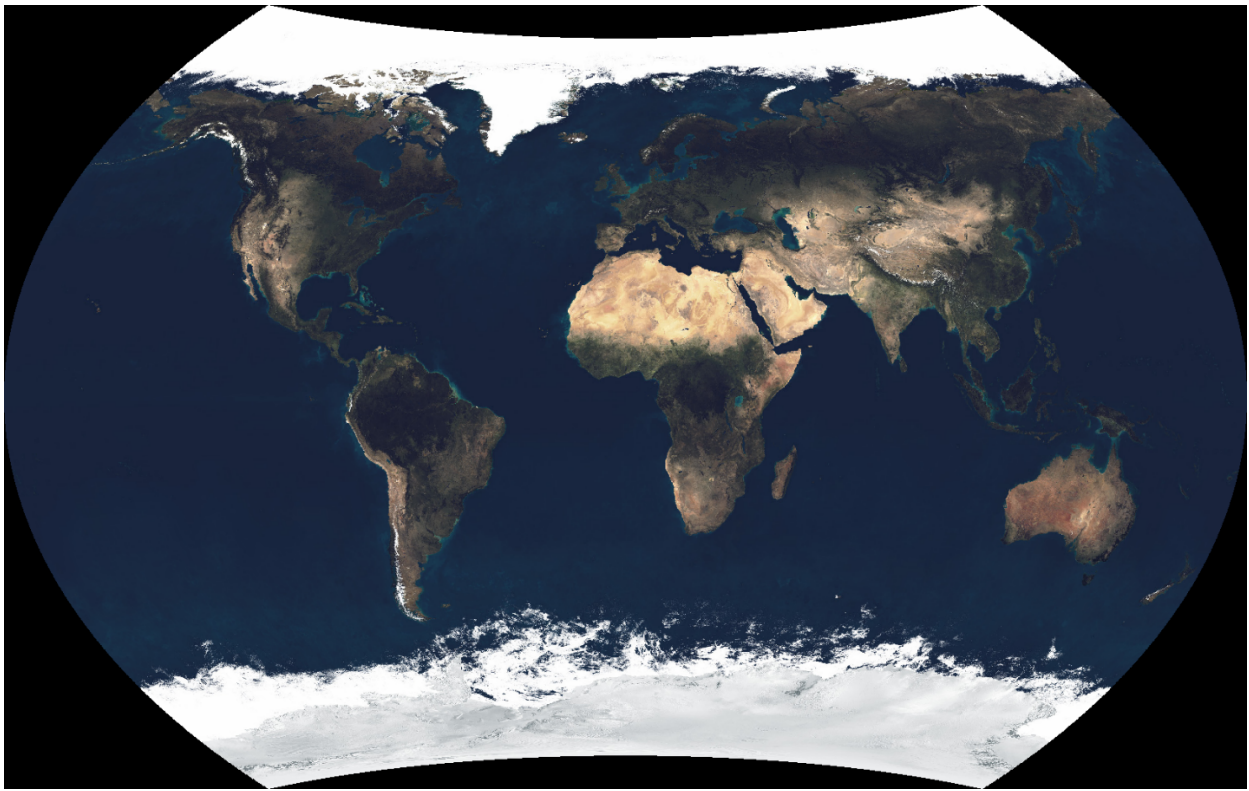

Figure 1. Winkel Tripel—National Geographic Version (Top) and Gott-Wagner (Bottom) Projections. NGS Winkel-Tripel: Open Source World Maps, Gott-Wagner: *Sizing up the Universe* (Gott & Vanderbei, 2010).

# The Gott-Wagner Projection

Map projections (like the Winkel Tripel) with linear scales along both the prime meridian and the equator do unusually well according to our fidelity metric. The Hammer-Wagner projection with a curved pole line was the best equal-area projection with $\Sigma\ \sigma^2 = 5.785$. A curved pole line seems beneficial as it lowers flexion. Thus, an Aitoff-Wagner projection, with a curved pole line but which is linear on both the equator and prime meridian might do well also. The Aitoff-Wagner projection multiplies each latitude on the globe by 7/9ths. The North Pole becomes the 70° latitude line; halve all longitudes and project the resulting hemisphere onto the plane with the azimuthal equidistant. Finally, multiply x- and y- coordinates by 18/5 and 9/7 respectively. It does not beat the Winkel-Tripel in terms of the six-term fidelity metric, but we can do better by using this Aitoff-Wagner as our starting point and adjusting via the coordinate mapping:

$x = 2\ z\ \cos(2\varphi/3)\sin(\lambda/2\sqrt{3})/\sin(z)$

$y = z\ \sin(2\varphi/3)/\sin(z)$

where $z = \arccos[\cos(2\varphi/3)\cos(\lambda/2\sqrt{3})]$ and $\varphi$ is latitude and $\lambda$ is longitude. This "Gott-Wagener" starts with the globe, multiplies latitudes by 6/9ths, longitudes by $(3)^{1/2}/4$, then maps onto the plane with the azimuthal equidistant, finally multiplying the x coordinate by 2. Parameters were chosen so as to have a (slightly curved) pole line a bit longer than the Winkel-Tripel to lessen flexion in longitude lines, and leave Africa no more squashed than it appears on the Kavraskiy-VII, which is acceptable. The goal was to get a low $\Sigma\ \sigma^2$ score while not "teaching to the test," so only this one projection was tried. It did beat the Winkel-Tripel with a score of $\Sigma\ \sigma^2 = 4.497$. It has better shapes and less flexion than the Winkel-Tripel while also having longitude plotted on a linear scale along the equator and latitude plotted on a linear scale on the prime meridian. It makes an excellent map of the Earth (see Figure 1). Australia in particular has a better shape than on the Winkel Tripel. This map was presented in *Sizing up the Universe* by Gott & Vanderbei (2010), where the Earth and Moon were shown to scale using this projection.

One might wonder why the best scoring maps discussed above were not round, as might be expected by symmetry considerations for a 2D representation of a sphere. Many old atlases featured two circular maps of the two hemispheres appearing side-by-side. Locally, these projections do quite well, but the boundary penalty, B=0.5, (corresponding to a cut of 360°) is twice that of the Equirectangular. A normalized error of 2 for the boundary term, squared is 4 which puts it out of the running for overall maps. And, interestingly, such side-by-side hemisphere maps are not popular for world atlases now. Cartographers have long known that more boundary cuts—interruptions—are a penalty, also inducing greater distance errors, which must be traded off against better local shapes and areas. It is a bit curious that the best overall flat maps of the sphere are not circular. It's a clue that we may be missing something. (And, as we shall ultimately see, the flat two-sided maps that we will be proposing in this paper which radically improve the errors, are indeed circular.)

## Standard Flat Maps--Topology Error

All the standard map projections have incorrect topology. An infinite plane has the topology of a sphere *minus* the south pole (as illustrated by the stereographic projection) . A finite map with a boundary, either at a pole or along a longitude line, marking the edges of the finite map, can be topologically distorted to cover a hemisphere of the sphere--but not the whole sphere. This is why many projections begin by cutting on the 180° longitude line, then halving the longitudes so as to map the sphere onto a hemisphere. The cut of course makes the topology different. The hemisphere can then be mapped onto the plane with say a Lambert equal-area projection to produce a circular equal-area map, which is then stretched by a factor of two in the x direction to make a Hammer equal-area projection.

Most world maps show the entire globe including the north and south poles in a finite region which can be printed on a page in a book. These maps are topologically incorrect, as they are topologically equivalent to a hemisphere rather than an entire sphere. Maps like the Winkel Tripel which stretch the poles into line segments also make an additional mistake as the North Pole is a point not a line.

## Folding and Unfolding as Map Projection Tools

We seem to be reaching a limit on improving the Winkel Tripel. It looks like only small improvements are in the offing. When that occurs in science, one often needs a breakthrough, some out-of-the-box thinking, to make any radical progress. Richard Feynman once said that in physics when we are stuck, when all the old methods don't work, then the new trick, the new method that is going to work is going to look very different than anything we have seen before.

Euclid allowed one to have a straight edge and a compass as tools. A perpendicular bisector of a line connecting two points could be constructed, an angle could be bisected, and many other things. But there was no way to trisect an arbitrary angle. Archimedes showed that if one was allowed to make two marks on the straight edge, an arbitrary angle could be trisected. Break the rules a little, add a new tool, and a previous roadblock could be overcome.

But in 1980, Hisashi Abe solved the angle trisection problem by using *fewer* tools. Give back the compass; give back the straight-edge. Now he could solve the trisection problem with the pen alone. He used origami! For a discussion, proofs, and axiomatic basis see Fuchs (2011).

The power of folding can be illustrated by an even simpler example. Suppose you want to draw the perpendicular bisector between two points. Following Euclid, you would (1) draw a large circle around the first point with radius larger than the separation between the two points,

(2) draw, using the same radius, a circle around the second point, (3) the circles will intersect in two points, connect these by drawing a line with the straight edge. Using origami, you can do it in two steps without either a straight edge or compass. Bend the paper in a U shape and place the two points on top of each other. Now crease the paper. The crease is the perpendicular bisector. Two operations instead of three and no tools! The angle trisection takes 7 steps. The cube root of 2 can be found in 4 steps. (See Fuchs, 2011).

The lesson here is that sometimes it is beneficial to try a radically different approach.

Map projections already use folding and unfolding as tools. Polyhedral projections like the Gnomonic Cahill Projection, and the Buckminster Fuller projection already produce flat maps that are unfolded versions of 3D polyhedral globes to approximate the sphere. Conversely, these flat maps may be folded up into polyhedral globes, healing their boundary cuts, as we discuss below.

We will be suggesting a new kind of flat map using the principle of folding.

## Polyhedral Globes, Envelope Polyhedra, and the Guyou Projection

The genesis of the idea for this paper comes from the paper "Envelope Polyhedra" (Gott, 2019). Five regular polyhedra are well known: the tetrahedron, cube, octahedron, dodecahedron, and icosahedron. All have equivalent regular polygons as faces and vertices with identical arrangements of polygons around them. They have less than 360° around a vertex, a delta function of positive Gaussian curvature at each vertex, and the same topology as a sphere. They obey Euler's equation for compact polyhedra: $V - E + F = 2$. They can be used, via the Gnomonic projection for example, to make 3D polyhedral globe maps of the Earth.

The Cahill butterfly map (see Snyder, 1993) can be folded up into an octahedron. There are some-delta function flexions and skewness on the edges with this polyhedral globe (which produce finite additions to the overall flexion and skewness errors), but the boundary cuts are eliminated and the distances are better. As a flat, unfolded map, the large boundary cut of 450° (four 90° cuts in the Southern Hemisphere from the South Pole to the equator, and one 90° cut from the equator to the North Pole) puts it out of the running for best overall map.

As shown in Goldberg and Gott (2008), boundary cuts B and skewness S are required in the error budget to avoid terrible maps winning the overall errors contest (daisy patterns with N separate gores as petals as N goes to infinity, or connecting these petals with webbing like webbed feet). These maps, which no one would choose for an overall best map of the world, cause either the boundary cuts or the skewness respectively to blow up.

For the celestial sphere, project it from a light at its center onto a tangent cube, via the gnomonic projection. This makes a celestial cube with the stars on the six interior faces, which can be unfolded, and then refolded, inside out to produce a celestial cube with the six square charts on the outside (Gott & Vanderbei, 2011, 48-79). North Circumpolar Stars appear on the cube's top face, four seasonal charts on the four sides, and South Circumpolar Stars on the bottom. Maps based on this design date back to 1674, notably including *Six Maps of the Stars* (Baldwin, Cradock, and Newton, 1831). Unfolded into a cross-shaped pattern to lie flat on the plane, the boundary cuts are 7x70.53° = 493.7°--worse than the Cahill.

When Buckminster Fuller presented his Dymaxion Earth map first (in *Life* in 1943) he claimed that his map was better than any previous map. He once posed with it *assembled* as a 3D polyhedral globe (a cuboctahedron). But the map, as presented in a magazine, cut open and splayed out on one plane surface, has boundary cuts of 660° losing easily to the Winkel Tripel on this account. Assembled, as a 3D crude globe, its local shapes and areas are good, it has no boundary cut error and its distances are greatly improved. Thus, Fuller was concealing its greatest flaw as a flat map. He was apparently aware of the boundary cut error, for he later oriented the cuboctahedron in such a way as to minimize the boundary cuts across land areas. While this orientation was good for the continents, it was terrible for oceans, ripping them apart into pieces on the page. But this was to be cured by making a model out of it--he included instructions on how to cut it out and fold it, with tabs which could be used to glue it together into its 3D shape. It was then not a flat map but a *3D polyhedron model* of a spherical globe. This had errors of course but they were less. The more faces the better. And he later adopted the icosahedron, the regular polyhedron with the most faces. But of course, one can make polyhedra with many more faces that would be still better. No one knew this better than he, for he had invented the geodesic dome. But he did not present a map with hundreds of faces, because it would have looked too interrupted on the page, and too hard to assemble.

Polyhedral globes have delta functions of positive Gaussian curvature at each vertex which integrated over (the infinitesimal area of) the vertex equal the angle deficit (360°- sum-of-polygon-angles) around the vertex. The total Gaussian curvature integrated over the (infinitesimal) area of all the vertices is $4\pi$, the same as the integrated Gaussian curvature integrated over a sphere. A cube has an angle deficit of 90° at each vertex (3 squares around each vertex rather than four). There are 8 vertices, so the total is 720°= $4\pi$.

Kepler noted the three regular planar tessellations (triangles—6-around-a-vertex, squares—4-around-a-vertex, or hexagons—3-around-a-vertex) were also regular polyhedra, but with an infinite number of faces. They had a sum-of-angles of 360° around each vertex and zero Gaussian curvature at each vertex.

Petrie and Coxeter (Coxeter, 1937), Gott (1967), and Wells (1977) discovered regular infinite sponge-like polyhedra (some with dihedral angles that were not equal). All had negative Gaussian curvature at each vertex (a saddle-shaped pattern of polygons) and a sum of face angles at each vertex which were greater than 360°. All had an infinite number of faces and were multiply connected (sponge-like) with an infinite genus.

These later played an interesting role in astronomy. Gott, Melott, and Dickinson (1986) showed that the theory of inflation implied that the initial fluctuations in density in the universe should have a sponge-like topology, with interlocking high- and low-density regions. They showed that under the action of gravity large-scale fluctuations would retain their sponge-like topology, leading to a structure today of clusters of galaxies connected by filaments of galaxies and low-density voids connected by low-density tunnels, a structure confirmed by many surveys and now called the *Cosmic Web* (see Gott, 2015). The Cosmic Web is well illustrated in Gott and Jurić's conformal, logarithmic Map of the Universe (Gott, Jurić, et. al. 2003, 2005).

In "Envelope Polyhedra" (Gott, 2019), new regular polyhedra were introduced where all vertices have identical arrangements of polygons around them but some of the dihedral angles are zero, allowing polygons to appear back-to-back. A rich variety of structures were found (over four dozen). Some were finite, some were infinite, some had positive, negative, or zero Gaussian curvature. Some of the finite ones are illustrated in the stereo pair in Figure 2.

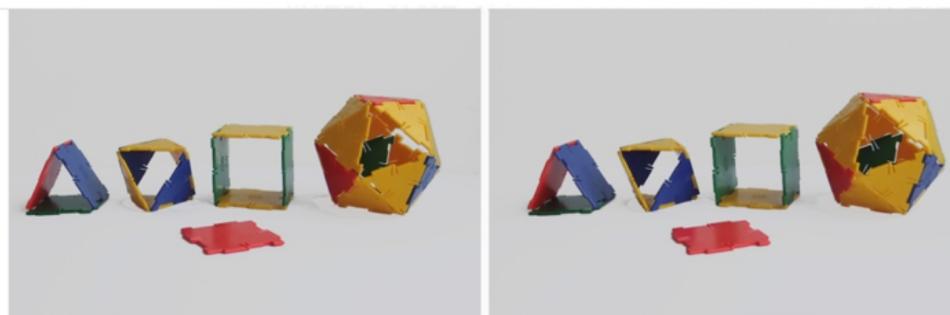

Figure 2. Some finite Regular Envelope Polyhedra (in stereo, from Gott, 2019). Of particular importance for this paper is the dihedron: squares—2-around-a-vertex, shown in red at the bottom.

From left to right at top are squares--4 around-a-vertex, triangles--6 around-a-vertex, squares--4 around-a-vertex, and triangles--8 around-a-vertex. The first three have the topology of a torus. The last, is a hollow icosahedron with four triangles missing, creating holes that link 16 exterior faces with 16 interior faces. This is a surface with negative curvature at each of its vertices and the topology of a sphere with three handles. Second from the right at top (squares--4 around-a-vertex) is a cube, missing its top and bottom tipped over toward us. It has 4 exterior faces and 4 interior faces. It has 4 interior edges, 4 exterior edges, and 8 edges linking the exterior and interior, for a total of 16. It has 8 vertices: F - E + V = 8 - 16 + 8 = 0 = 2(1- genus), giving a genus of 1, a torus topology. This particular envelope polyhedron is discussed in (Gott and Vanderbei, 2020): "A Simple Isometric Embedding of the Square Flat Torus in 3D space."

At bottom is squares—2-around-a-vertex. This has two faces, one on top and one on the bottom. It is like an unopened envelope. It has 4 vertices, and 4 edges. It is a convex polyhedron of zero volume. It satisfies Euler's equation for convex polyhedra, F - E + V = 2, i.e. 2 - 4 + 4 = 0. It therefore has a genus of zero and the topology of a sphere. In squares—2-around-a-point, Gaussian curvature is concentrated in four delta functions at the four vertices with integrated value at each equal to the angle deficit of 180° = π. The general case is N-gons—2-around-a-point, with 2 faces, N vertices, N edges, giving F - E + V = 2 - N + N = 2. These all have the topology of a sphere. These *dihedra* had been known before (see Coxeter,1937). These are also regular polyhedra.

Of particular relevance is the conformal world map projection of Guyou of 1886, shown in Snyder (1993). A Western hemisphere is mapped onto a square, at left, which is attached to an Eastern hemisphere mapped onto a square at right. This makes a 2x1 rectangular map of the entire Earth (Figure 3). The boundary cut is thus only 270° instead of the 360° one would have if the hemispheres were depicted as circles. The two squares are joined like conjoined twins by a median line bisecting the map. This line follows a longitude line and covers an arc of 90°--from 45°N latitude to 45°S latitude. We are now going to apply folding.

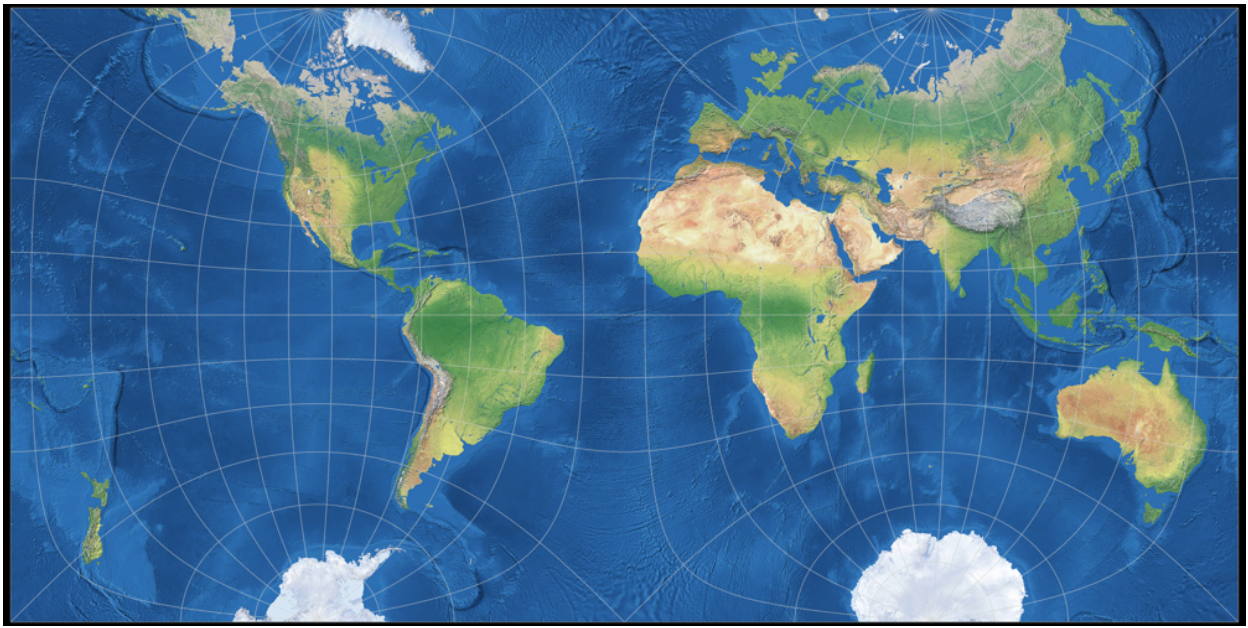

Figure 3. Conformal World Map of Guyou 1886. Photo Credit: Daniel R. Strebe

This is the central idea for Gott's new type of double-sided map: We are going to fold along this median vertical line, so that the Eastern Hemisphere folds backward like folding a billfold to join up with the Western Hemisphere at all its edges. Now tape the edges together.

You now have a *square envelope* with the Western Hemisphere on the front and the Eastern Hemisphere on the back. This is the envelope polyhedron, squares—2-around-a-vertex shown in Figure 2 . Importantly, it has the topology of a sphere. This map has 0 boundary cut error. The original map had a normalized boundary cut error of 1.5. Distances are also greatly improved. We measure distances on the map by stretching a string between points A and B on the map. If A is in the Eastern Hemisphere and B is in the Western Hemisphere, the string will pass over an edge. Two points on opposite sides of the boundary cut are close on the globe, but can be at opposite sides of the original Guyou map: the fractional distance error blows up as such points move closer to each other on the globe. But on the envelope this problem does not occur.

To conclude, the squares—2-around-a-vertex assembled Guyou map is a flattened model of the sphere. *It belongs to the set of polyhedral models of a sphere*. But it is also a *flat map*. It has zero volume and can be printed on the front and back of a single sheet of paper or plastic. *It is a map that belongs in two categories. It is the first polyhedral globe we have encountered that is flat (of zero volume), and it is the first flat map we have considered that is double-sided with the correct topology of a sphere.*

An octagonal conformal map of a hemisphere is more accurate than a square one, and a folded map, octagons—2-around-a-vertex would also be more accurate than a folded Guyou projection. Consider the dihedra N-gons—2-around-a-vertex in the limit where N goes to infinity. This would, Gott realized, give in the limit even more accurate back-to-back *circular* maps. Each vertex of a double-sided N-gon has an angle deficit of $4\pi/N$, giving a total integrated curvature of $4\pi$—like a sphere. In the limit as N goes to infinity this total integrated curvature is banished to the equatorial edge of the disk, where by symmetry arguments it is made harmless against producing additional flexion or skewness errors at the edge. Consider a geodesic crossing the equator: by symmetry, on the map, as long as the same projection is used for both sides, the angle it makes with the equator (or edge) is the same on both sides, so there is zero kink or bend in the geodesic as it crosses the equator. Likewise, the velocity of the map image of a truck traveling on that geodesic does not change as it crosses the equatorial edge of the map. Thus, there are no delta function flexion or skewness errors associated with the equator. One can imagine an ant circling the outer edge of the map traveling straight and at constant speed, so it is a geodesic on the globe that has no flexion or bending on the map. The equator itself is a set of measure zero with no blow up of local errors in isotropy or area so it contributes nothing to these errors as well. The two-sided map has the topology of a sphere, deforming into a sphere in a simple way: as a rubber circular envelope which you inflate from the inside into a lens shaped pillow and then into a sphere.

## Map projections for double-sided disks

We propose two-sided azimuthal polar projections for whole world flat maps. The North Pole is at the center of the top circular face and the South Pole is at the center of the bottom face, straight longitude lines fan out uniformly from the poles and a single function (radius as a function of latitude) suffices to describe the projection. We get indications of how to determine good radial functions from existing one-sided projections.

Chebyshev showed the best conformal map of any closed boundary region of the sphere (in the sense of minimizing the sum of the squares of the scale errors) is one that has constant scale along the boundary (see Snyder, 1993, page 140). According to this criterion the best conformal map of a hemisphere is a circular disk made using the stereographic projection. This beats any polygonal conformal map of the hemisphere.

This projection has a remarkable property: pairs of antipodal points on the map are all equidistant. Let the disk radius be r = 1. Random point A is a distance $r_A$ from the center of the top of the disk. Its antipodal point A', by symmetry, is a distance $r_{A'} = r_A$ from the center of the bottom face of the disk, along a diameter of the disk, but pointed in the opposite direction. The shortest distance from A to A' on the map starts at A, goes outward along a radial line a distance $(1 - r_A)$ to reach the circumference of the disk, then on the bottom of the disk inward along a radial line a distance of 1 to get to the South Pole, and then a distance $r_{A'}$ outward along the same diameter to get to point A'. The total distance on the map traveled from A to B is $(1 - r_A) + 1 + r_{A'} = (1 - r_A) + 1 + r_A = 2$ for any pair of antipodal points on the map. This is true for any projection where $r_A$ is a monotonic function of latitude; thus, it is true for a Lambert azimuthal equal area projection and an azimuthal equidistant projection as well as for a stereographic projection. This lowers distance errors for points widely separated on the globe.

More generally, calculating the distance (length of the shortest path) between a random point $(r_A, \varphi_A)$ in the Northern hemisphere and different random point $(r_B, \varphi_B)$ in the southern hemisphere on the map (where $\varphi_A$ and $\varphi_B$ are the longitudes) is equivalent to Alhazan's billiard problem (Dorrie, 1965): finding the point on the circular edge of a circular billiard table where one should aim the cue ball to bounce once off the edge and hit another ball at an arbitrary point on the table. Ptolemy formulated the problem in 150 AD, and Alhazen (who lived 965-1040 AD) discussed it in his book on optics. It is unsolvable with straightedge and compass. Its solution in Cartesian coordinates requires solution of a cubic equation (e.g., Reide 1989, Neumann 1998), found by Huygens (1895). But it is solvable using folding (origami) (Alperin, 2004). Interestingly, our map projection involves this famous and deep mathematical problem which circles back to origami. When we calculate the distance between two points on the map, we solve this numerically.

Below in Table 2 are the isotropy, area, flexion, skewness, distances, and boundary cut errors for different projections. Also listed are the normalized sum of squares of errors—where the Gott double-sided disk projections are much better.

## Table 2: Comparison of Projections

| σ | I | A | F | S | D | B |
|---|---|---|---|---|---|---|
| Equirectangular | 0.51 | 0.41 | 0.64 | 0.60 | 0.449 | 0.25 |
| Winkel-Tripel | 0.49 | 0.22 | 0.74 | 0.34 | 0.374 | 0.25 |
| Gott-Wagener | 0.395 | 0.319 | 0.682 | 0.367 | 0.397 | 0.25 |
| **Double-Sided Disk projections** | | | | | | |
| Stereographic | 0 | 0.39 | 0.37 | 0.37 | 0.135 | 0 |
| Lambert | 0.36 | 0 | 0.52 | 0.11 | 0.098 | 0 |
| Equidistant | 0.239 | 0.129 | 0.462 | 0.062 | 0.086 | 0 |

**Some Standard Projections--Normalized squared errors ($\sigma^2$)**

| $\sigma^2$ | I | A | F | S | D | B | $\Sigma\sigma^2$ |
|---|---|---|---|---|---|---|---|
| Equirectangular | 1 | 1 | 1 | 1 | 1 | 1 | 6 |
| Winkel-Tripel | 0.923 | 0.288 | 1.337 | 0.321 | 0.694 | 1 | 4.563 |
| Gott-Wagener | 0.600 | 0.605 | 1.136 | 0.374 | 0.782 | 1 | 4.497 |
| **Double-Sided Disk Projections** | | | | | | | |
| Stereographic | 0 | 0.903 | 0.336 | 0.384 | 0.091 | 0 | 1.714 |
| Lambert | 0.71 | 0 | 0.81 | 0.18 | 0.218 | 0 | 1.240 |
| Equidistant | 0.221 | 0.096 | 0.518 | 0.01 | 0.036 | 0 | 0.881 |

The (Azimuthal) Equidistant projection with a linear radial scale scores best of all (lowest $\Sigma\sigma^2$).

If θ is π/2 minus the latitude then the radius of a latitude circle is r(θ) = θ. The next simplest relation would add a quadratic term of amplitude α using the functional form:

$$r(\theta) = \frac{2\theta}{\pi} + \alpha \frac{2\theta}{\pi}\left(1 - \frac{2\theta}{\pi}\right)$$

for the Northern hemisphere (and similar for the Southern). Since all polar projections can be approximated by a polynomial expansion in θ, we note that a Lambert projection is well approximated by $\alpha = 0.16$ while the Stereographic is approximately $\alpha = -0.35$. The Equidistant projection corresponds to $\alpha = 0$. We have experimented with this quadratic model to try, numerically, to identify an optimal radial function. See Figure 4.

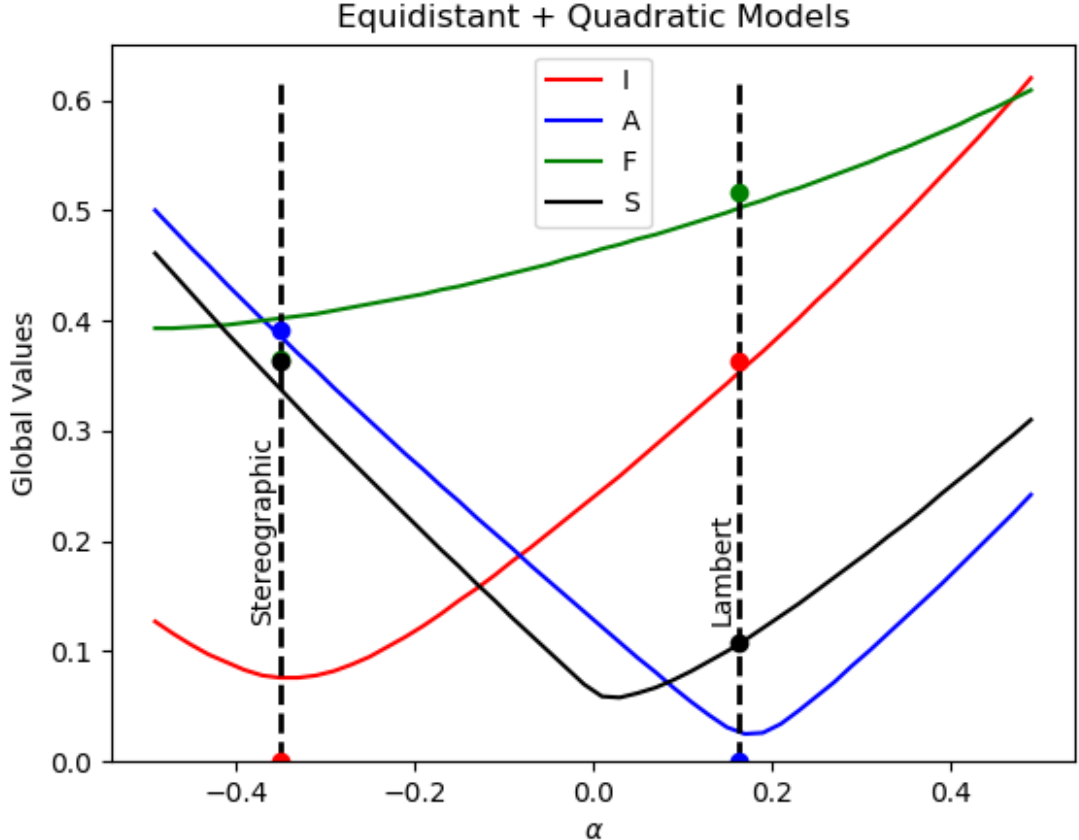

Figure 4. Equidistant + Quadratic Models. Errors in Isotropy, Area, Flexion, and Skewness, as a function of the amplitude and sign of the quadratic term. The Equidistant model for r(θ) has α = 0 and is linear, while positive α is concave downward like the Lambert, and negative α is concave upward like the Stereographic. Dashed lines indicate the best fits for the quadratic term to the Lambert and Stereographic. Colored dots indicate the actual error values for the Lambert and the Stereographic.

    The colored dots are close to, but not exactly on the curves because the quadratic models do not perfectly model the Lambert and Stereographic. But the values of the curves at α = 0 are exact for the Equidistant. Interestingly, Skewness has a sharp V shaped minimum very near α = 0. This, together with the fact that Isotropy and Area have error functions with opposite slopes near α = 0, and Flexion is gently rising there, means that the minimum sum of squares of errors solution is likely to lie very close to α = 0. This suggests that, avoiding teaching to the test, which would depend on the exact weighting factors applied to each error, the Equidistant Azimuthal is the simplest and best solution. It has an additional nice property: uniform scale on all meridian lines.

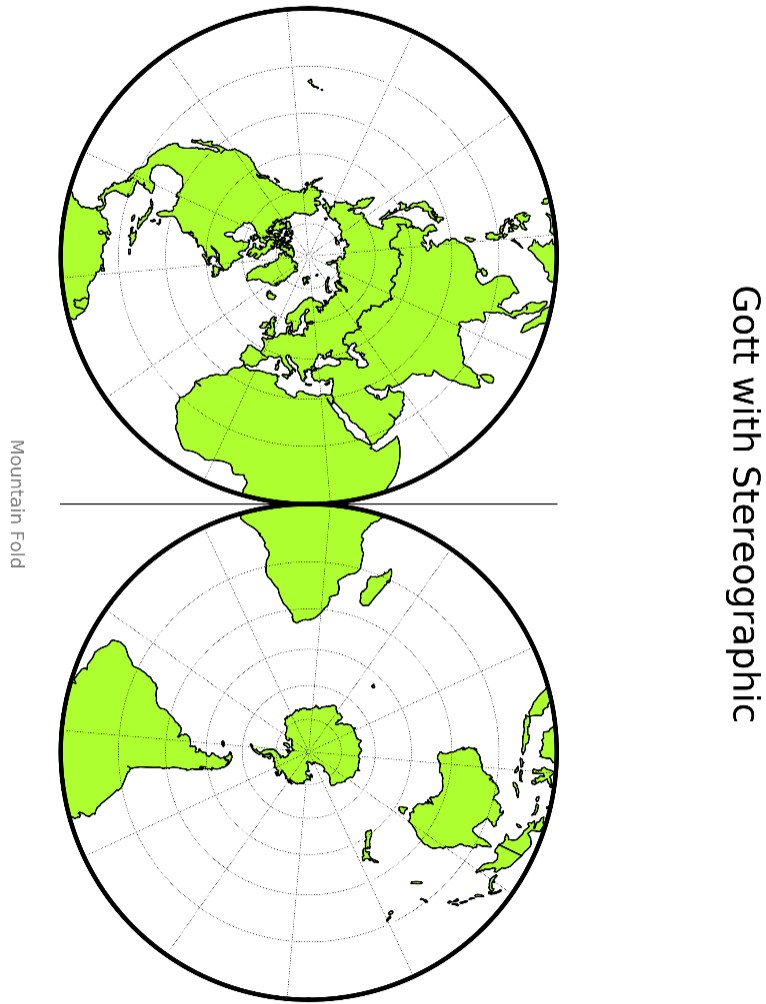

Figure 5. Double-sided Gott Stereographic. Make a mountain fold along the horizontal line. Glue the Northern and Southern Hemispheres together back-to-back (using double stick tape for example) and cut out, to make a double-sided disk (like a phonograph record). It's conformal (local shapes are perfect here--zero isotropy error), but areas at the edge are 4 times enlarged relative to those in the center. Antarctica looks too small relative to Australia for example.

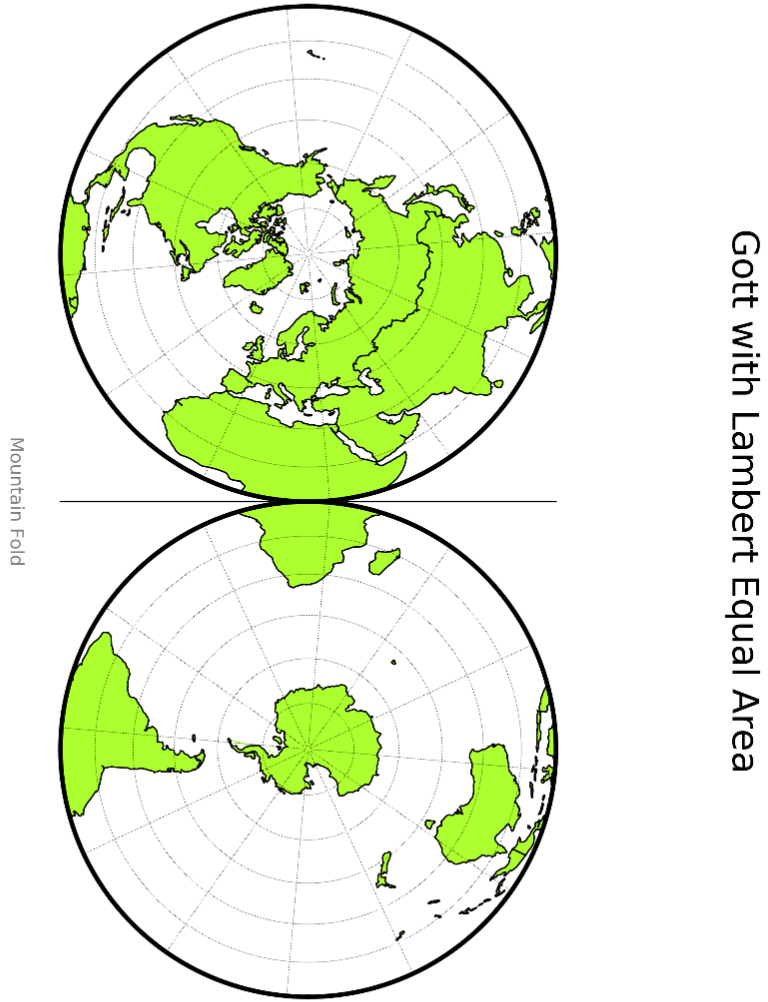

Figure 6. Double-sided Gott Equal Area. Fold, glue back-to-back, and cut out as in Figure 5. Areas are perfect but shapes are distorted locally 2:1 at the edge. Australia looks too squashed.

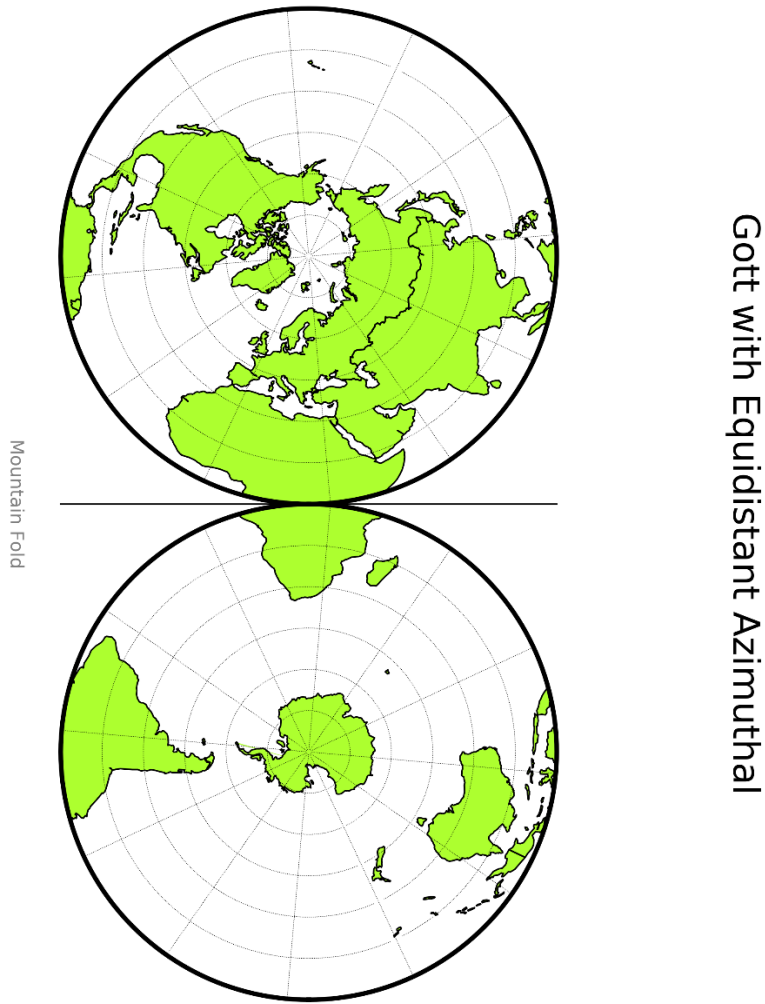

Figure 7. Double-sided Gott Equidistant Azimuthal. Uniform scale along meridians. A compromise projection like the Winkel Tripel. Errors in isotropy and area are less noticeable. Areas at edge are $\pi/2$ larger than at center. Local shapes are $\pi/2:1$ elongated at the edge. Antarctica does not look too small, and Australia does not look too squashed. In sums of squares of normalized errors in isotropy, area, flexion, skewness, distances and boundary cuts, this is the best overall flat map yet produced.

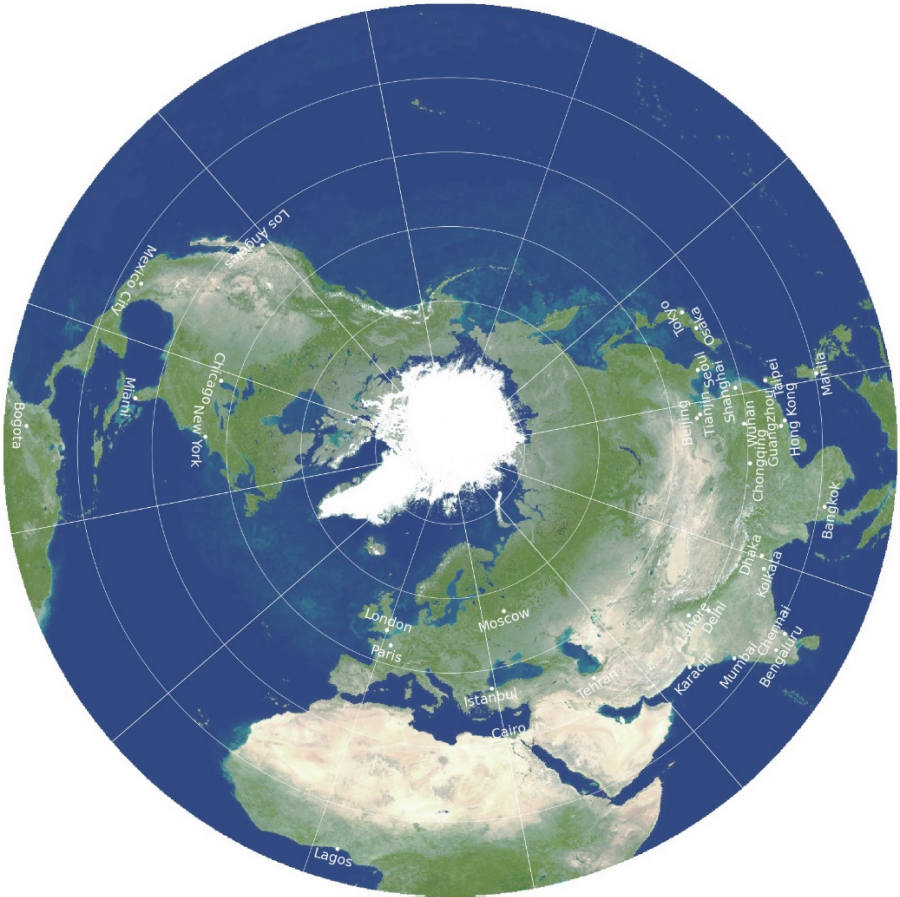
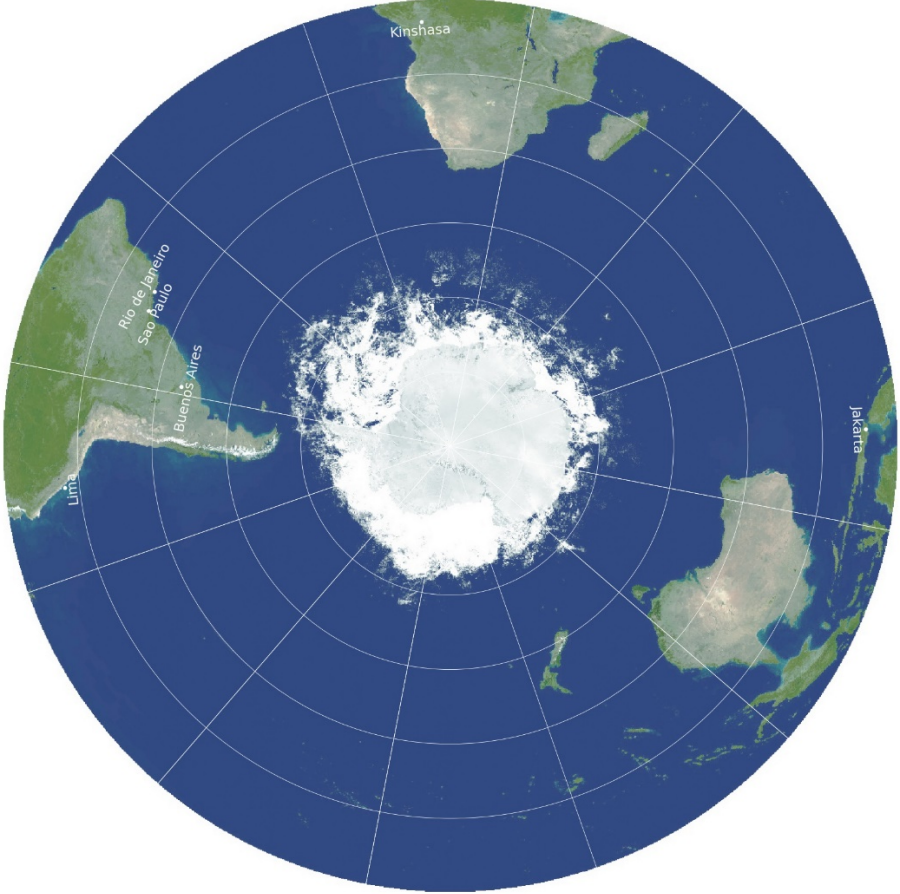

*Figure 8* The least error Gott double-sided disk, Azimuthal Equidistant projection from Figure 7 presented in larger format with labeled cities. With the hemispheres folded back-to-back and taped together, the map is continuous across the equator. This is, we believe, the most accurate flat map of the Earth yet. Uses adapted NASA imagery. This movie shows how it looks when assembled. https://vanderbei.princeton.edu/tmp/Earth_gif.mp4. For a larger version that can be printed back-to-back, ready for cutting out, see the Appendix.

Needless to say, this Double-Sided Gott Equidistant projection (Figure 8) has radically improved sums of squares of errors relative to the Winkel Tripel. Errors in isotropy, area, flexion, and skewness are lower in the double-sided map because it is easier to map one hemisphere at a time.

It has better distances because the topology is correct and there are no boundary cuts. Points close on the globe are a short string distance away from each other on the map. The Gott double-sided equidistant map beats the Winkel Tripel map in all 6 error measures! Thus, it is superior regardless of normalization procedures.

Distance errors between points in Figure 8 are bounded by ±22.2%. By contrast, errors in distance in the Winkel Tripel blow up both as one approaches the North and South Poles and for pairs of points close to each other on opposite sides of the boundary cut in the Pacific. The Equidistant version (Figure 8) is superior in this regard to either the conformal or the equal area versions (Figures 5 and 6), which have distance errors bounded by ±33.3%. In fact, the Equidistant version (Figure 8) is optimal for this distance bound for any double-sided disk projection. In Figure 8, the upper and lower limits are found for points close to each other on the equator on the high end and points 180° apart on the low end—with ratio $\pi/2:1$. Since the scale across the diameter of the Figure 8 map is uniform along a diameter of the map, the lower limit obtains throughout for pairs of points on the diameter, but if the scale is not uniform the lower limit will be lowered further for some pair of points. The double sided Guyou Map has distances that blow up in the corners, as does any double-sided N-gon map. Figure 8 produces the lowest distance errors of the three double-sided maps illustrated, with $D = 0.086$, decisively beating the Gott-Mugnolo, the one-sided circular map with the best distance error of $D = 0.341$ (Gott, Mugnolo, and Colley, 2007).

Figure 8 is a flat map with vastly improved errors compared with the Winkel Tripel. This is a flattened globe; but, being flat, this map has a volume of zero and is much more convenient to store.

You could print a Mercator map on the circumference of a cylinder to eliminate the 180° boundary cut and improve distances. Japan is now closer to Hawaii, as it should be, rather than on the opposite side of a wall map. But this cylinder would still not be as good as the maps we are proposing because our combined isotropy, area, flexion, and skewness errors are lower. Cylinder maps have the property that you can observe only half of the map of the Earth at one

time--just like the real globe. When Thomas Edison invented the phonograph, he recorded the wiggly lines to produce the sound on cylinders. These took up volume. These were supplanted in a short time by double-sided phonograph records (invented by Emile Berliner): circular disks. They take up almost zero space relative to the cylinders. You can have a whole stack of phonograph records.

We are proposing something similar, a flat circular map that utilizes both the front and the back and can be stacked. It can be made out of paper and inserted loose leaf into your *National Geographic* magazine. It could be made out of a thin disk of cardboard, or even better out of a thin disk of plastic, giving some rigidity. A thin box could hold flat double-sided maps of all the major objects in the solar system. Or there could be a stack of Earth maps giving physical data, political boundaries, population density, climate, languages, explorers' voyages, empires at different historical epochs, continents at different geological epochs, and so forth. Many interesting box sets could be imagined. Long playing phonograph records probably tell us the most convenient size: 12 inches in diameter. For Earth this would mean 1,036.4 miles to the inch along all meridians. In this paper, for comparison with other projections, we always assume spherical geometry. (Note, however, that this projection could easily be adapted to use the ellipsoidal geometry of Earth. Radius in the map would be distance along the ellipsoid from the North Pole, in which case the scale along all meridians would be 1,035.8 miles to the inch.)

Has anyone ever made maps like this before? Snyder's *Flattening the Earth: Two Thousand Years of Map Projections* (1993), a compendium of nearly 200 map projections does not include any, nor does Mark Monmonier's (2018) list of map projections that have received patents. We have found at least one child's double-sided jig-saw puzzle, which was circular and which had the Eastern Hemisphere on one side and the Western Hemisphere on the other side. But these were more drawings than maps, in particular since Antarctica appeared on only one side of the map. There was no intention here to draw a more accurate map. Also, puzzle pieces are thick, which is not wanted. In addition, a number of solar system mobiles have planets as paper disks, printed on both sides. But these are really just pictures, as shown by the fact that Saturn also has its rings printed on both sides. Often, the same face of the planet is shown on both sides. By contrast, we have proposed double-sided disk maps of the globe with the explicit intention to minimize the important mathematical errors of isotropy, area, flexion, skewness, distances and boundary cuts. Our stereographic version beats all one-sided conformal maps, our Lambert equal-area version beats all one-sided equal-area maps, and our equidistant version is best of all flat maps for minimizing the six error terms, making it the most accurate map of Earth yet.

A disadvantage of our map ( Figure 8) is that you can't see all of the map at once. This is often cited as an advantage of flat maps, in addition to their zero volume. Should an error for this be applied to our map? Should it be part of the error budget for maps? In the Goldberg and Gott (2008) paper the six errors considered are errors made in depicting, on a flat map, the spherical surface of the globe: the smaller the sum of squares of errors, the higher the verisimilitude (fidelity) to the actual spherical Earth. Of course, the globe itself, by this token, must have zero errors. It is a perfect map of itself. *And the globe can't be seen all at once, only half at most*. So, in that sense, *our double-sided flat map is actually more like the globe of the Earth than the other flat maps*. Thus, it would not seem proper to introduce such an error term. Also, as we have commented before, our map has the same topology as the globe and wastes no paper surface. Finally, it is not even quite true that you can't see all of our map at once. That is because you have two eyes! Tip the disk vertically with its equatorial edge aligned with your nose. Your left eye can see the entire Northern Hemisphere and your right eye can see the entire Southern hemisphere. It appears as splayed outward like butterfly wings, because of the parallax angle between your stereo eye views. You can use this trick when inspecting the continuity of South America and Africa as they drape across the equator.

      For Earth maps, we suggest the most accurate azimuthal equidistant map projection be used in the polar projection (as in Figure 8). Simple to understand, it is like the United Nation flag (which extends into the southern hemisphere but not to include Antarctica.) Longitude lines are straight lines, and latitude lines are circles. The equator, of fundamental importance, is at the disk's rim. North Pole and South Poles, special points, are located at the centers of the front and back. Most of the most populous cities in the world are located in the Northern Hemisphere which is on one side. Since the Cold War, as possible missile trajectories arc over the North Pole, the advantage of having the North Polar areas and relationships displayed prominently has been emphasized, particularly by Briesemeister. As climate change continues, polar shipping lanes will become important, as polar flight paths already are. It might be complained that when you hold up the map to look at Eurasia, the United States is "upside down." But this is "northern bias." Besides, since you are holding this in your hand, you can easily rotate it around to view the United States "right side up."

      There may be, however, some political reasons for using an equal-area map. When Peters (1983) produced his equal area projection (a rediscovery of the Gall 1885 projection), he argued that the Mercator was culturally unfair to developing countries that were shown smaller than they should be. Thus, if one desired an equal-area projection, one could use the Lambert equal-area projection, shown in Figure 6. Its shapes are a bit worse at the equator 2:1 elongated, instead of π:2 elongated. This makes Indonesia look flattened. But it might be popular for political maps. The WinkelTripel is not equal area and it has not come in for much criticism however.

      Now, for some astronomical maps. It makes sense to have Eastern and Western Hemispheres for the Moon, with the familiar Earth-facing side of the Moon on one side of the map and the far side of the Moon on the other (Figure 9 at top). We use the equatorial aspect of the equidistant azimuthal projection. There is of course no change in the error budget. There are more dark mare (old lava flows of basalt rock) on the near side where the crust is thinner. Figure 9 (at bottom) is a map of the Earth made in the same way for comparison. Maps with this orientation might be appropriate for hanging from the ceiling with a string allowing them to rotate due to room air currents.

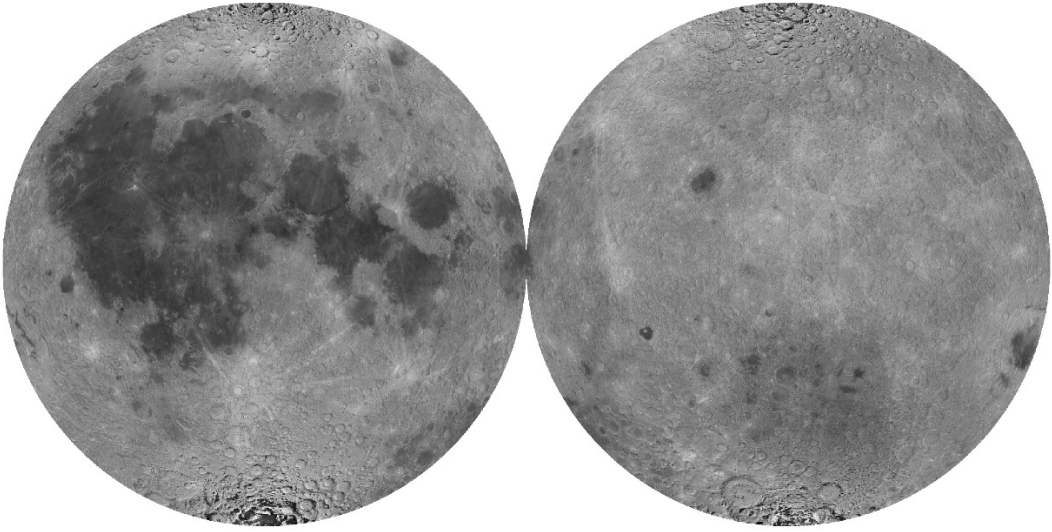

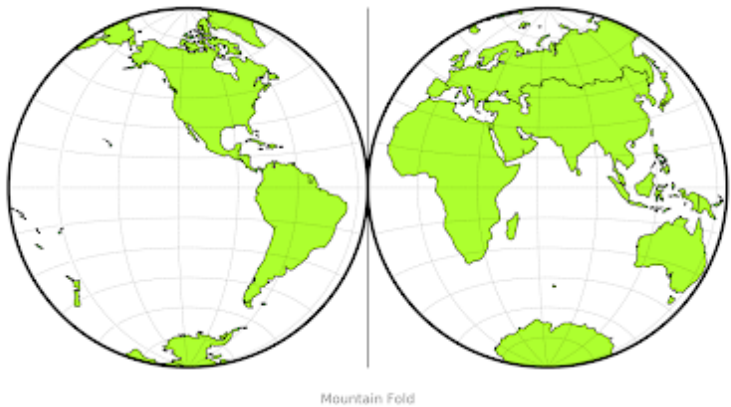

Figure 9. Near side (left) and far side (right) of the Moon (at top). Make a vertical mountain fold in the middle, tape front and back sides together back-to-back and cut out to make a disk (like a phonograph record). Earth shown in the same way (at bottom). These are equatorial aspect equidistant azimuthal projections.

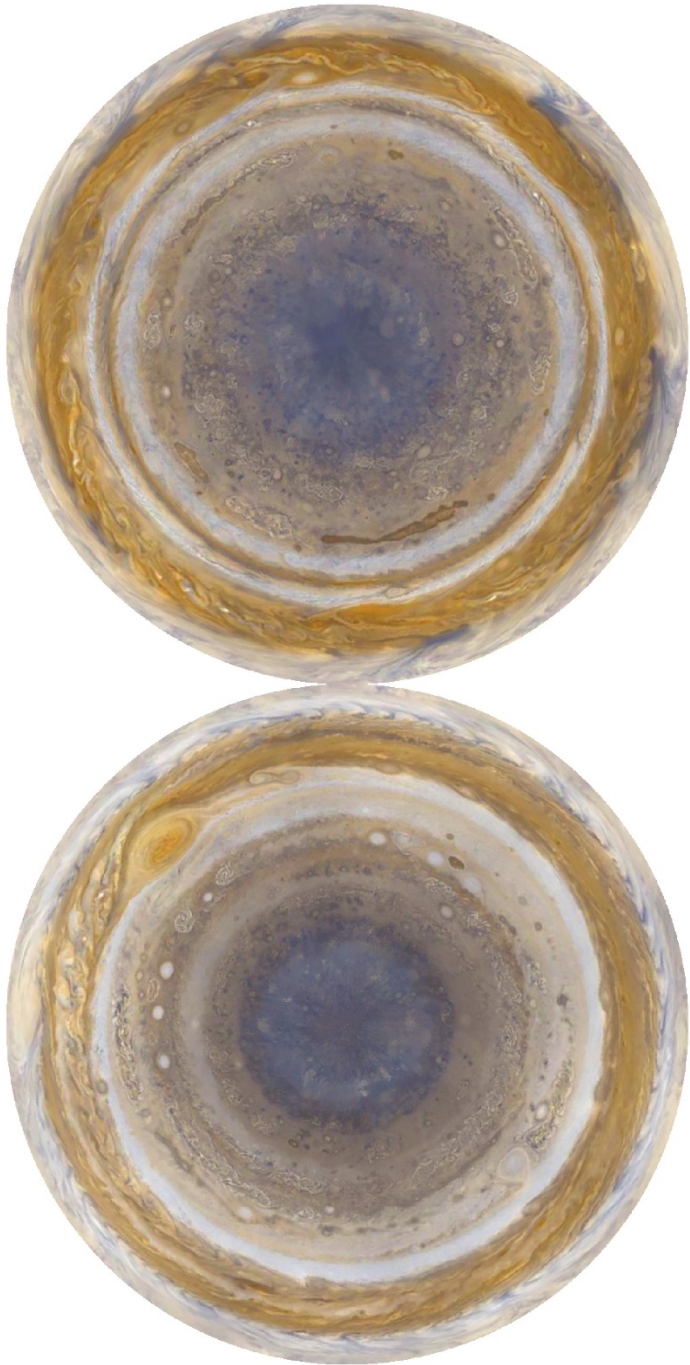

Figure 10. For Jupiter, this polar projection of the Gott Equidistant Azimuthal shows the cloud belts as circular as they follow latitude lines. The great Red Spot appears on the Southern side (bottom of the disk). In this and remaining pictures, make a horizontal mountain fold in the middle, glue the northern (top) and southern (bottom) hemispheres together back-to-back and cut out.

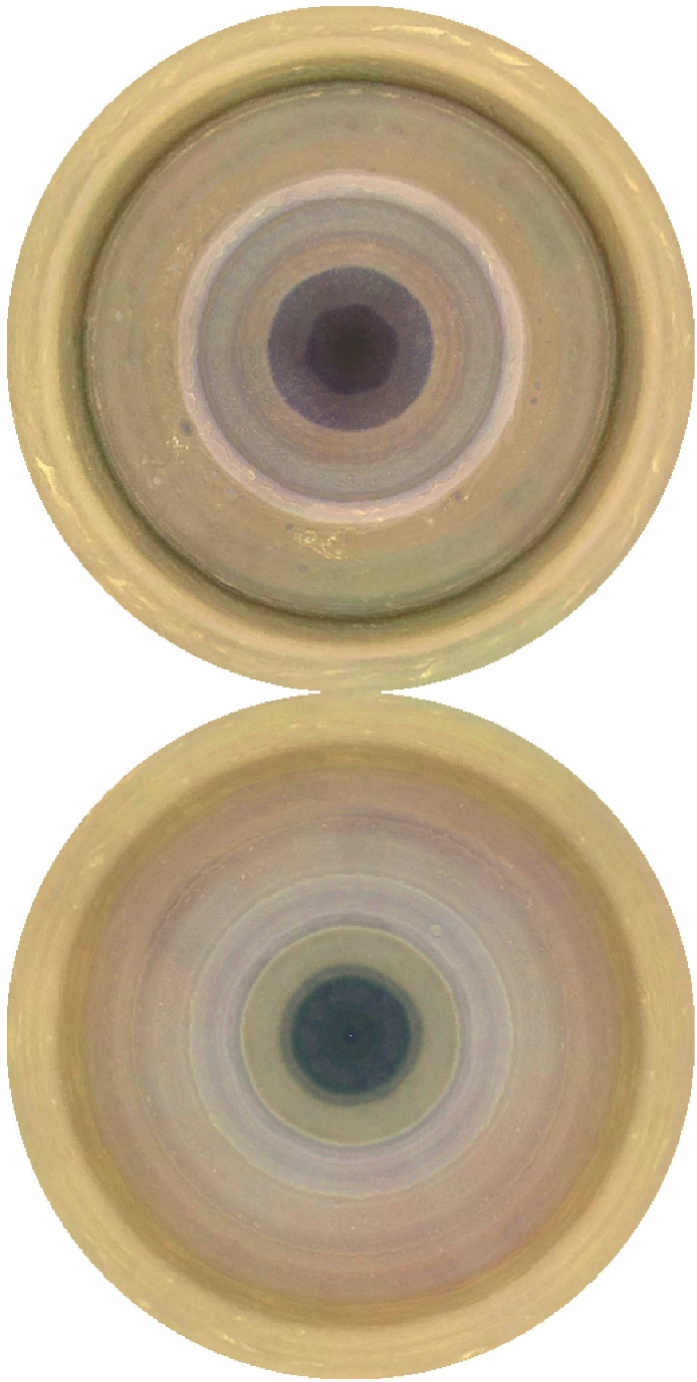

Figure 11. Double sided Gott Equidistant Azimuthal polar projection of Saturn shows its unusual hexagonal North polar storm beautifully.

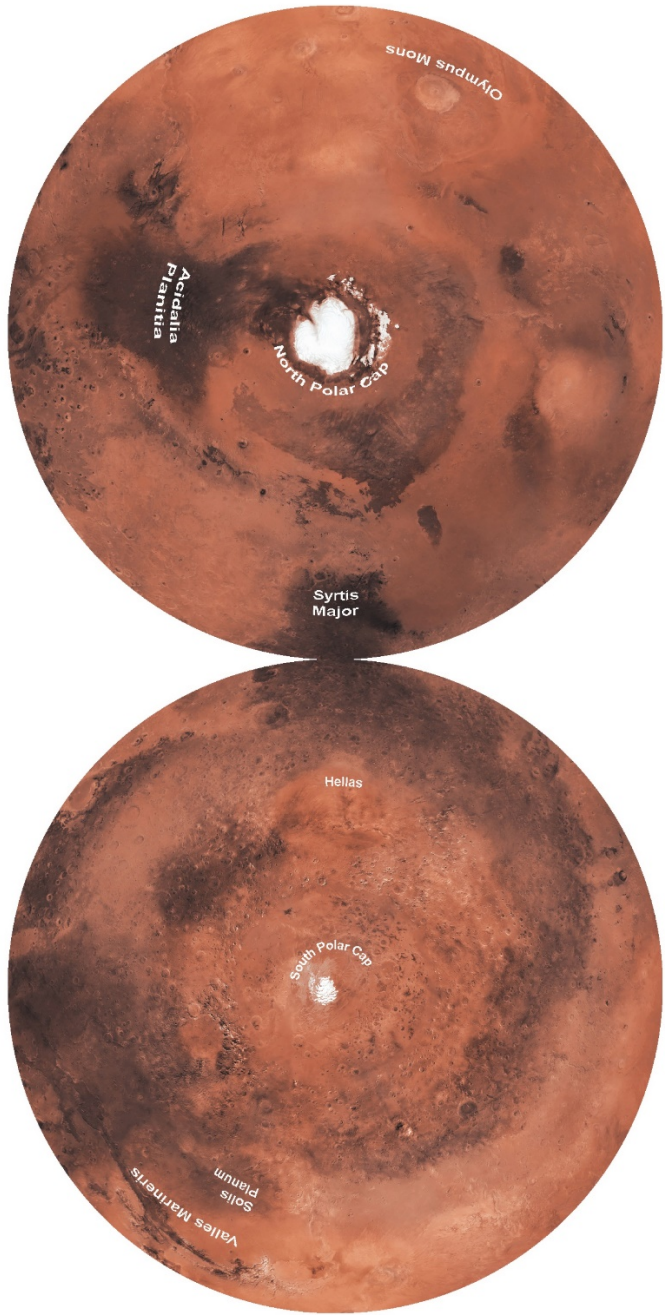

Figure 12. Mars, Gott double-sided Azimuthal Equidistant. Its large North Polar Cap made of water ice is well shown, as is Olympus Mons, a volcano 70,000 ft tall--Mars' highest mountain. Syrtis Major, a prominent dark region of basalt rock not covered by reddish sand and dust, crosses the equator from north to south at the center where the two hemispheres touch at the mountain fold. Hellas is an old impact basin filled with light-colored sand and dust. The South Polar Cap is at the center of the southern hemisphere. Valles Marineris is a giant canyon.

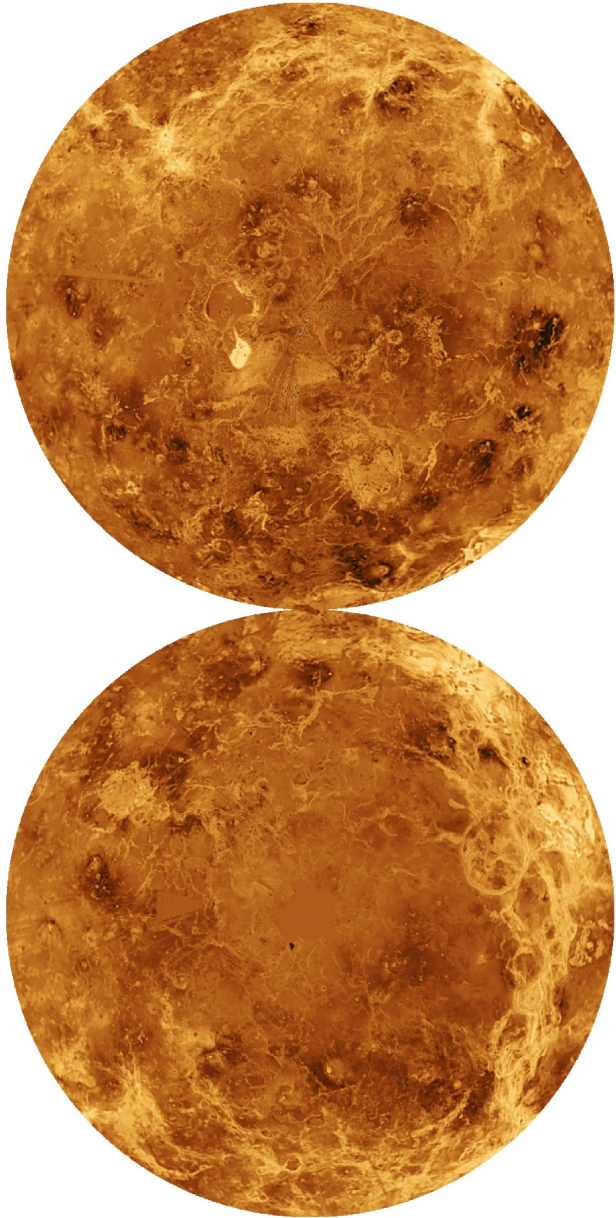

Figure 13. Venus surface map from radar observations.

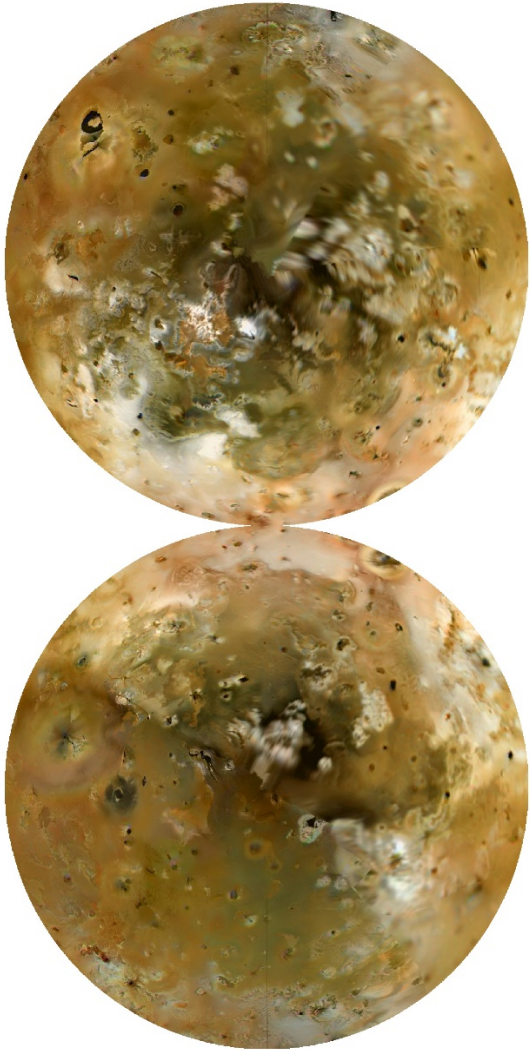

Figure 14. Jupiter's moon Io, pockmarked with volcanos.

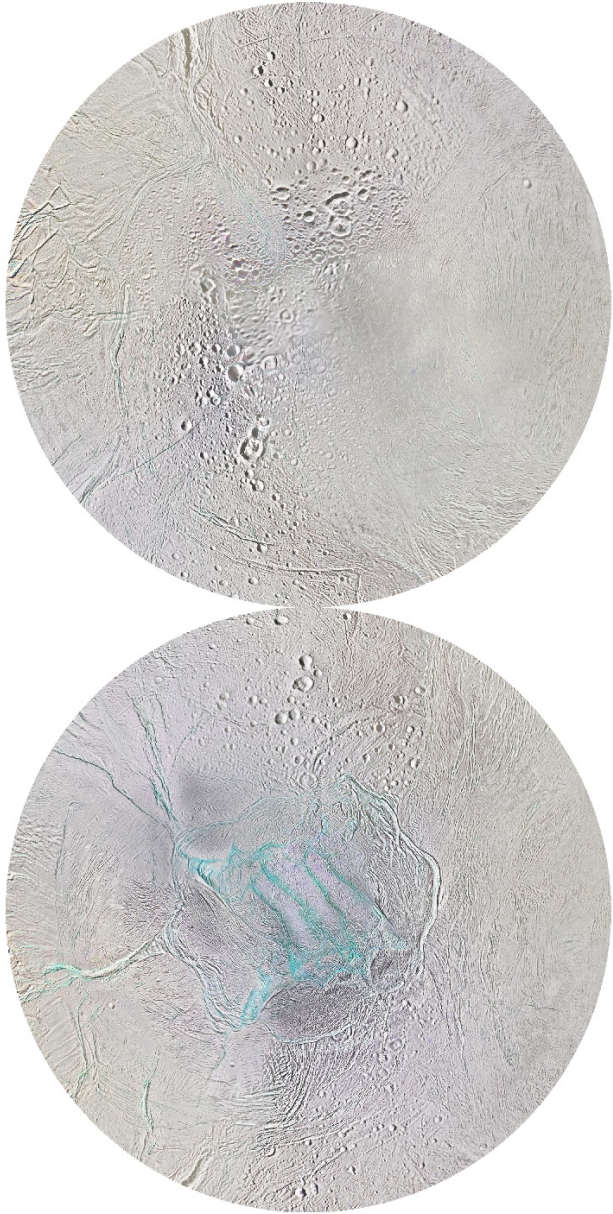

     15. Saturn's moon Enceladus on a double-sided Gott equidistant projection. The map beautifully shows in blue the parallel tiger stripes near the South Pole in the center of the bottom face. These are cracks in the surface ice over a southern ocean, from which geysers spew salty water vapor into space. Much of this ultimately falls back on the surface as snow. The northern hemisphere, lacking an ocean covered by surface ice constantly being resurfaced, is more cratered.

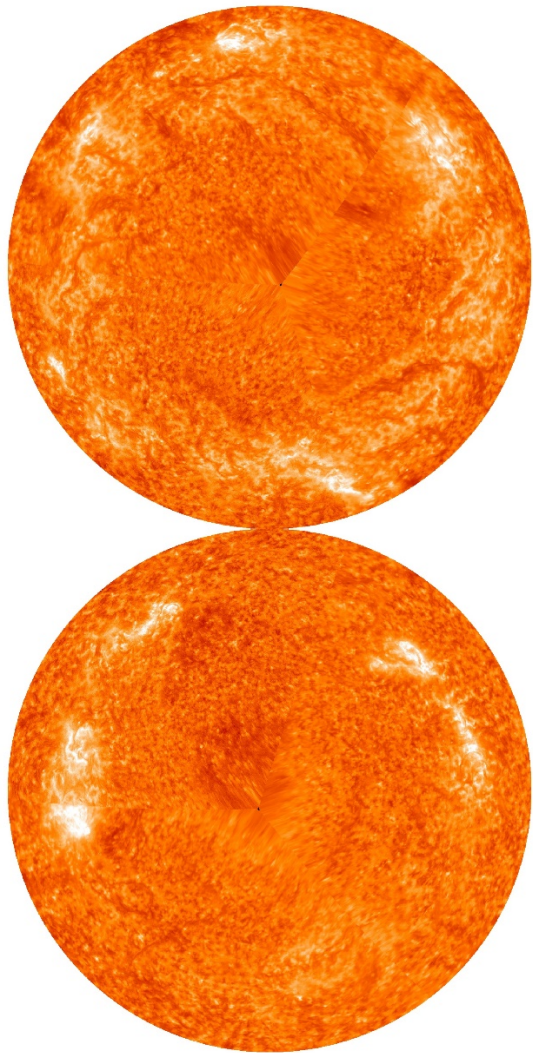

Figure 16. A double-sided Azimuthal Equidistant map of the Sun using photos taken by the two STEREO satellites on opposite sides of the Sun so that they provided for the first time full coverage of the Sun's entire surface at one moment. Taken in the extreme UV, it shows features in the Sun's chromosphere and inner corona, just above the visible surface seen in visible light. Data from NASA.

For maps of the celestial sphere, we definitely want a polar projection. And the equidistant projection is also preferred as it would not crush important constellations like Orion as much as the Lambert azimuthal. Stereographic maps produce too much distortion in area to be used frequently. National Geographic for example chose the equidistant azimuthal projection for its side by side Northern and Southern Sky Maps. Interestingly, Changbom Park (1996) has shown that the famous Korean star chart (Ch' on-Sang-Yul-Cha-Bun-Ya-Ji-Do) dating from 1675 also uses the polar equidistant azimuthal. One wants a Northern Hemisphere map on one side of the double-sided phonograph record map because this shows the North Circumpolar stars in the center, and can be rotated in your hand to follow the rotation of the sky.

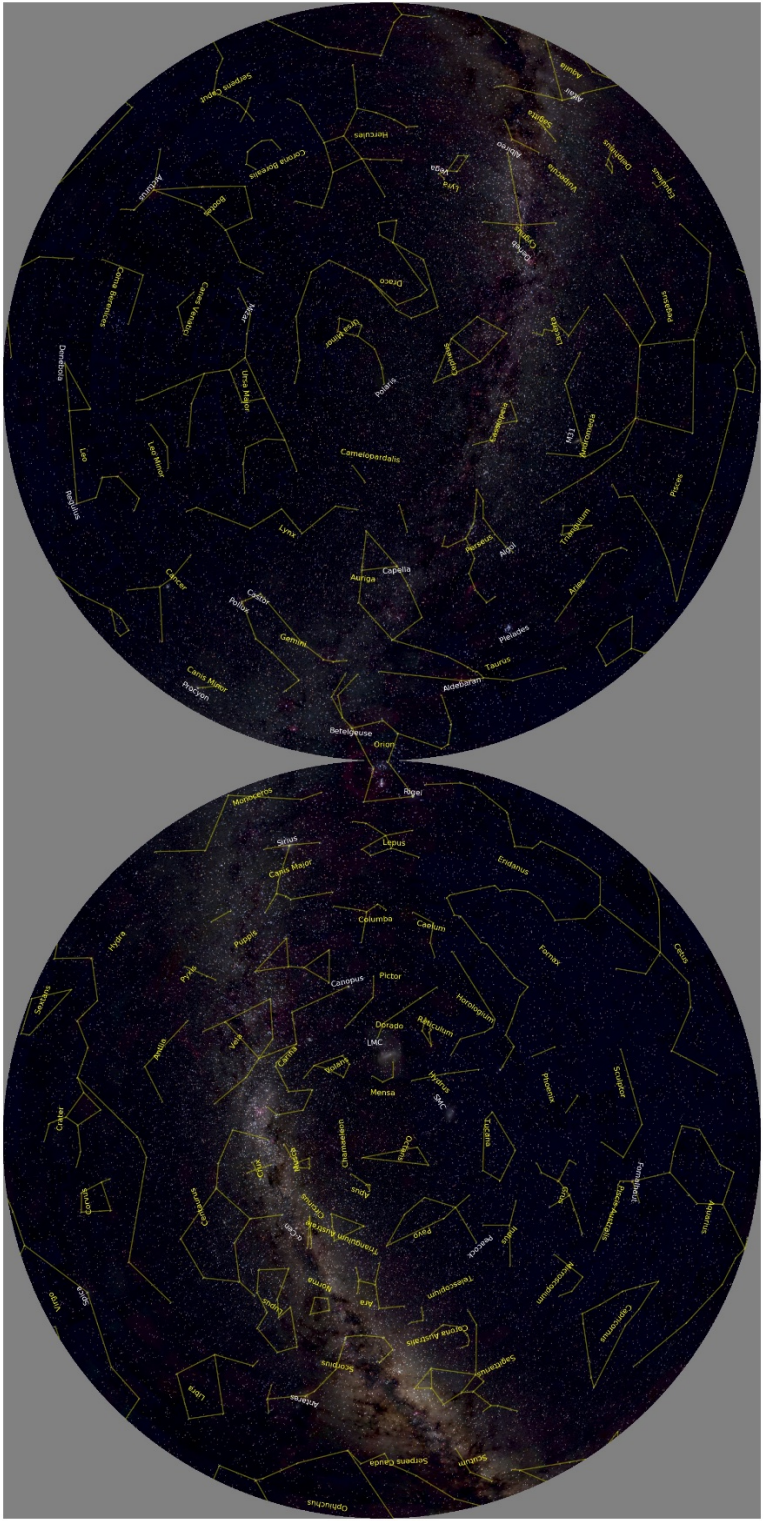

Figure 17. Double sided Gott Equidistant Azimuthal sky map. Northern and Southern Hemisphere stars. The Milky Way is more prominent in the south.

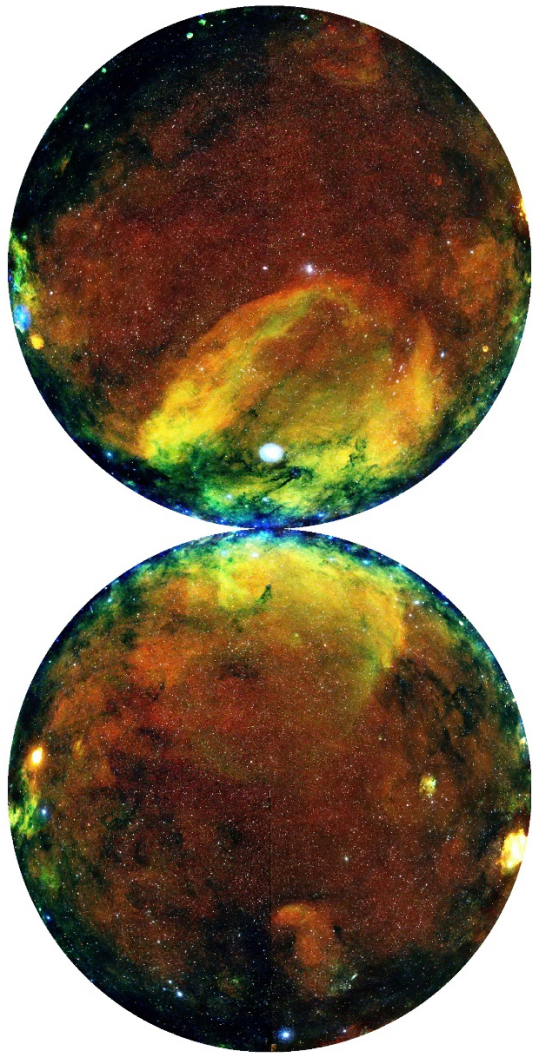

Figure 18. X-ray sky from eRosita (data from J. Sanders, H. Brinner & eSASS team (MPE)/E. Churazov, M. Gilfanov (on behalf of IKI). Red (300-600 electron volts [eV]), green (600-1,000 eV), and blue (1,000 -2,300 eV). The North Galactic Hemisphere is at top and the south galactic hemisphere is at bottom with a horizontal mountain fold in the middle where the two charts touch at the galactic center. The galactic equator is at the circumference of the double-sided disk. The large yellow region (green + red) in the north galactic hemisphere at top is a bubble of hot gas in the Milky Way. The brightest white point source in the north galactic hemisphere is Sco-X1, an X-ray source due to gas from a companion star in a binary star system falling on the surface of a neutron star. At the North Galactic Pole, in the center of the Northern Galactic Hemisphere is a small white dot, due to hot gas in the Coma cluster of galaxies. Near the galactic equator in the south are a yellow blob at 3 o'clock (the Vela supernova remnant, and a blue dot at 6 o'clock (the Crab Nebula pulsar).

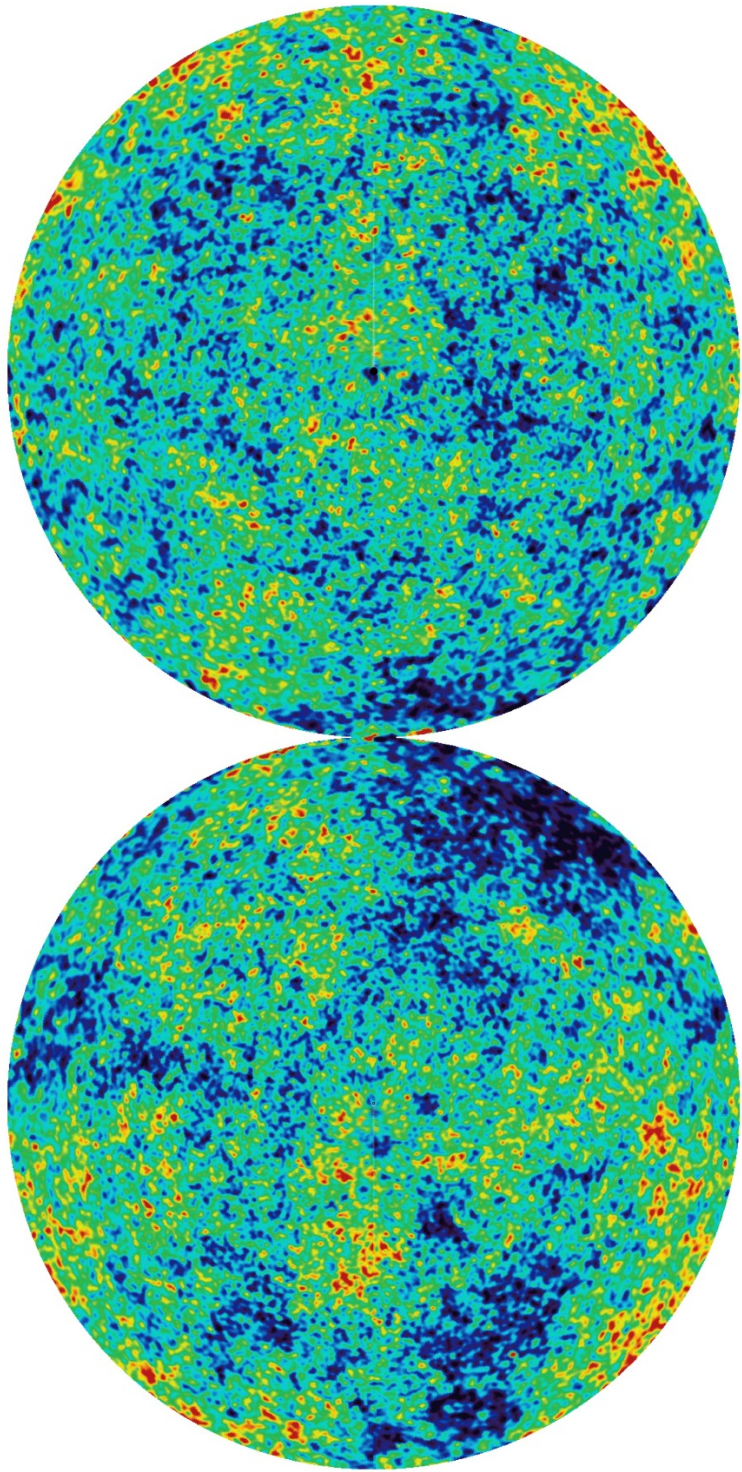

Figure 19. Cosmic Microwave background (NASA:WMAP), depicted as in figure 18. Temperature fluctuations of roughly 1/100,000 are shown with red (hottest) to blue (coolest).

In the end, double-sided disk maps would have, for the first time for flat maps, the correct topology of a sphere, and map errors radically improved from those in the Winkel Tripel. The Winkel Tripel is a map you can hang on your wall; these double-sided maps are more accurate ones you can hold in your hand.

Acknowledgements: We thank Chuck Allen for helpful comments.

## Appendix

To make a double-sided flat map, print the next two pages on both sides of a single sheet, flipping on the long edge. The Northern and Southern hemisphere views should perfectly align, front and back, on the page. Then, simply cut carefully on the circumference and you have a double-sided flat map of Earth.

Here is a gif movie showing how it will look:

https://vanderbei.princeton.edu/tmp/Earth_gif.mp4

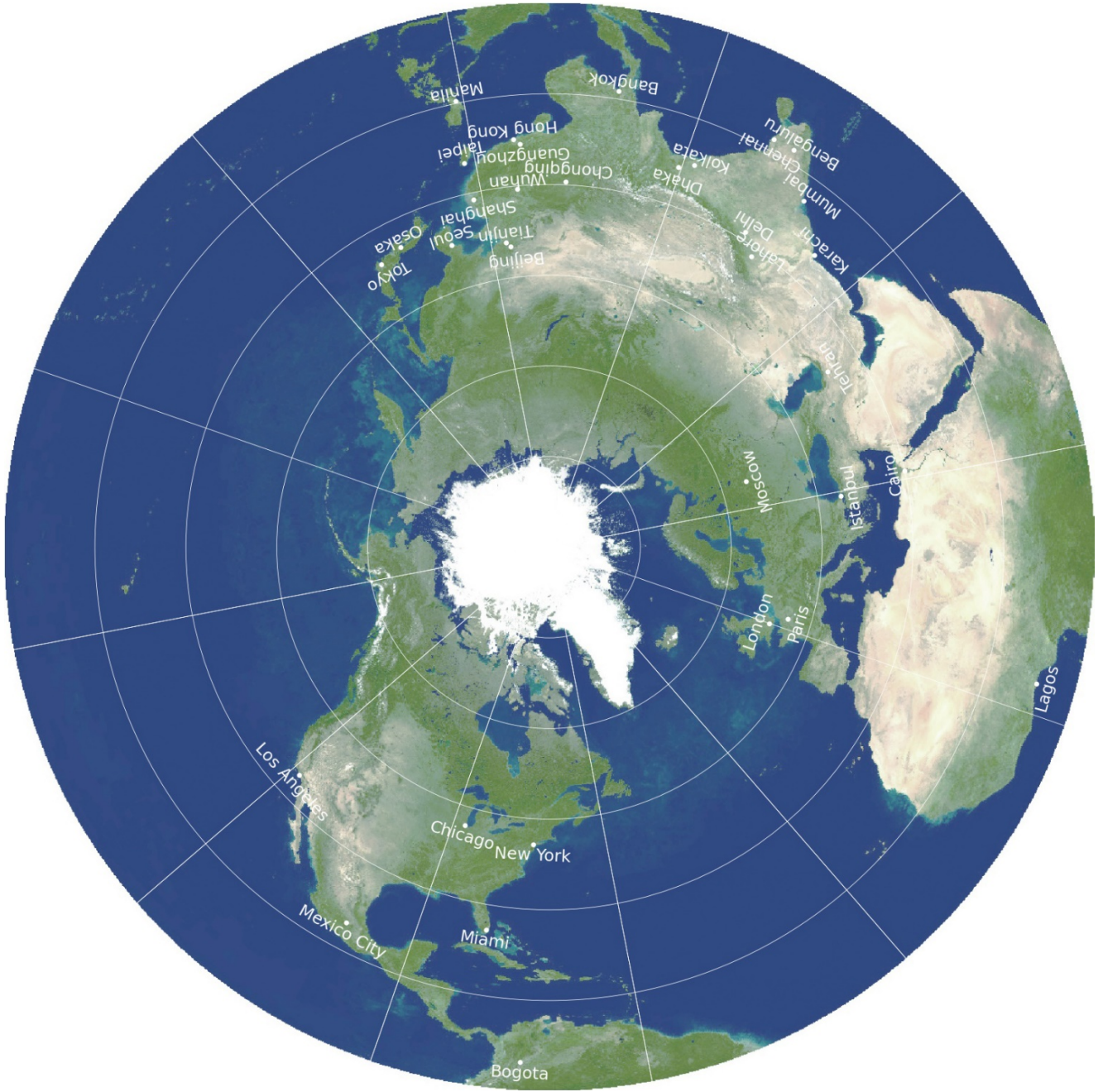

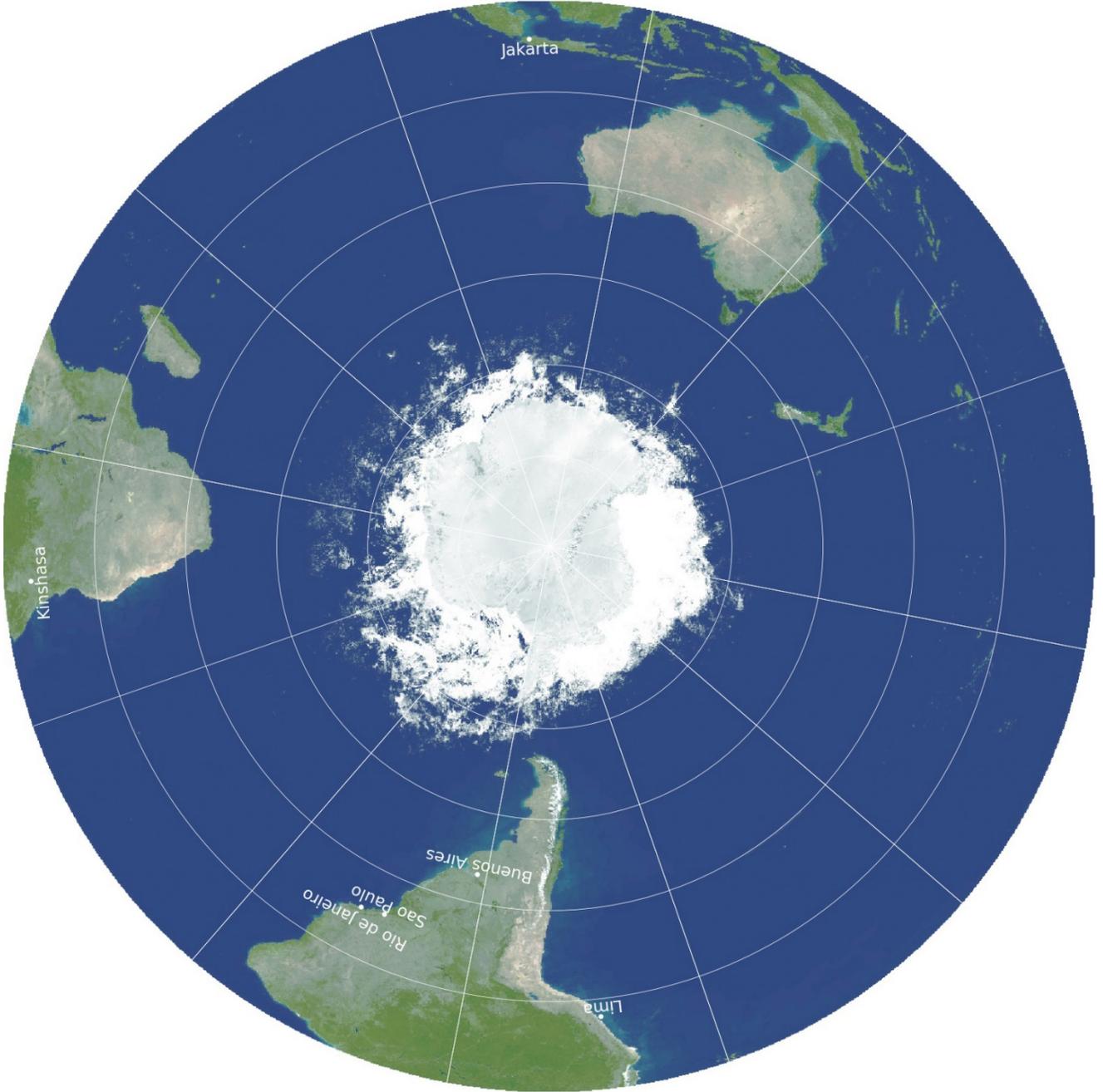

2